# A Robust Statistical method to Estimate the Intervention Effect with Longitudinal Data

Mohammad M. Islam, Ph.D. and Erik L. Heiny, Ph.D.
Department of Mathematics, Utah Valley University,
800 W University Parkway, UT-84058, USA

Segmented regression is a standard statistical procedure used to estimate the effect of a policy intervention on time series outcomes. This statistical method assumes the normality of the outcome variable, a large sample size, no autocorrelation in the observations, and a linear trend over time. Also, segmented regression is very sensitive to outliers. In a small sample study, if the outcome variable does not follow a Gaussian distribution, then using segmented regression to estimate the intervention effect leads to incorrect inferences.

To address the small sample problem and non-normality in the outcome variable, including outliers, we describe and develop a robust statistical method to estimate the policy intervention effect in a series of longitudinal data. A simulation study is conducted to demonstrate the effect of outliers and non-normality in the outcomes by calculating the power of the test statistics with the segmented regression and the proposed robust statistical methods. Moreover, since finding the sampling distribution of the proposed robust statistic is analytically difficult, we use a nonparametric bootstrap technique to study the properties of the sampling distribution and make statistical inferences. Simulation studies show that the proposed method has more power than the standard t-test used in segmented regression analysis under the non-normality error distribution. Finally, we use the developed technique to estimate the intervention effect of the Istanbul Declaration on illegal organ activities. The robust method detected more significant effects compared to the standard method and provided shorter confidence intervals.

Keywords: Interrupted time series analysis, Intervention effect, autocorrelation, robust median test

Correspondence to: Mohammad M. Islam, Email: mislam@uvu.edu

1. Introduction:

Interrupted time series is a quasi-experimental design used to evaluate the intervention effect in longitudinal data [1, 2]. Segmented regression analysis is a powerful statistical method to find the effects of implemented policy interventions on interrupted time series data [3]. In a segmented regression analysis, before and after the intervention, each segment of the time series is allowed to exhibit different levels and trends [1]. Segmented regression analysis can also allow researchers to include variables other than the intervention that can potentially cause a change in the level and/or trend of the outcome of interest [2].

Segmented regression analysis is suitable when the outcome variable is serially ordered as a time series and when several observations are available in both pre-intervention and post-intervention periods [1]. A change in the trend of the outcome after the policy intervention indicates the intervention effect. A detailed description of segmented regression analysis and its methodological guidance are found in [4, 5]. Different statistical approaches can be applied to estimate pre- and post-intervention levels and trends [1, 2].

The segmented regression model fits the least square regression line to each segment of the independent variable-time [1, 3]. The accuracy of the statistical results found by using the segmented regression model depends on four assumptions: linearity between time and the outcome variable, statistical independence (no autocorrelation) of the errors, homoscedasticity of the errors, and normality of the error distribution [1, 3, 6].

If any of these assumptions is violated, then the hypothesis testing, confidence intervals and forecasts developed by the segmented regression model are inefficient and biased. The presence of autocorrelation causes the standard errors of the estimates to be inflated, and consequently, the type II error rate increases [1, 3]. Linden [3] developed STATA commands to detect and control the autocorrelation based on NEWEY-WEST [14] and PRAIS [15].

The construction of confidence intervals and significance tests for the coefficients of the segmented regression model are based on the assumption of normally distributed errors. Confidence intervals may be either too wide or too narrow if the error distribution is not normal. In large sample studies, the normality assumption is robust, and valid results can be obtained using the normal approximation.

In a small sample study, if the errors are not normally distributed, then estimating the intervention effect using segmented regression analysis leads to incorrect inferences. The normality assumption is not robust in small sample studies, and the normal approximation is therefore not appropriate. To address the small sample problem and non-normality in the outcome variable, the intervention effect on longitudinal data is estimated using the rank-based method of Theil and Sen for a nonparametric slope. In addition to this, a bootstrap based testing procedure is developed for testing the significance of the intervention effect.

In this article, we investigate the effect of the violations of the normality assumption on the power of the standard t-test used in segmented regression. We compare the standard t-test with a robust rank-based median test for the slope of the line calculated by Theil and Sen's nonparametric approach and use a bootstrap based test to determine whether the intervention coefficient is significantly different from zero.

## 2. Statistical Model and Hypothesis Testing:

### 2.1 Intervention effect estimation:

Consider the standard linear Interrupted Time Series Analysis (ITSA) regression model

$$Y_t = \beta_0 + \beta_1 T_t + \beta_2 X_t + \beta_3 X_t T_t + \varepsilon_t, \tag{2.1}$$

where $Y_t$ is the outcome variable measured at equally-spaced time point $t$; $T_t$ is the time since the start of the study; $X_t$ is a dummy variable representing the intervention ( pre-intervention periods $X_t = 0$ , otherwise $X_t = 1$); $X_t T_t$ is an interaction term. The model (1) can be written as $Y_t = \beta_0 + \beta_1 T_t + \varepsilon_t$ for the pre-intervention outcomes, and $Y_t = (\beta_0 + \beta_2) + (\beta_1 + \beta_3)T_t + \varepsilon_t$ for the outcomes after the intervention. Therefore, $\beta_0$ is the initial level of the outcome variable and $\beta_1$ is the slope or trend of the outcome variable prior to the intervention; $\beta_2$ is the change in the level of the outcome between the pre- and post-intervention model; $\beta_3$ is the difference between the pre-intervention and post-intervention slopes or trend of the outcome variable.

Although the goal of segmented regression analysis is to evaluate whether there is a change in the level ($\beta_2$) or trend ($\beta_3$) of an outcome after an intervention, we focus on estimating the change in the slope (or trend) of the regression model before and after the intervention, $\beta_3$, and its corresponding inference. If the change in the slope before and after the intervention is zero, we can conclude that there is no effect due to the intervention on the trend of the outcome variable of interest.

The segmented regression model (1) can be written as $\boldsymbol{Y}_{n\times 1} = \boldsymbol{W}_{n\times 4}\boldsymbol{\beta}_{4\times 1} + \boldsymbol{\epsilon}_{n\times 1}$, where we assume that $\boldsymbol{\epsilon}_{n\times 1} \sim N_n(\boldsymbol{0}_n, \sigma^2 \boldsymbol{I}_n)$. The least squares estimate of $\boldsymbol{\beta}_{4\times 1}$ is

$$\widehat{\boldsymbol{\beta}}_{4\times 1} = (W'_{n\times 4} W_{n\times 4}) \boldsymbol{W'}_{4\times n} \boldsymbol{Y}_{n\times 1}, \tag{2.2}$$

where $\boldsymbol{W}_{n\times 4}$ is the design matrix.

Since $\widehat{\boldsymbol{\beta}}$ is a linear function of $Y$, $\widehat{\boldsymbol{\beta}} = (\boldsymbol{W'W})^{-1}\boldsymbol{W'Y} \sim N_4(\boldsymbol{\beta}, \sigma^2(\boldsymbol{W'W})^{-1})$ where $\boldsymbol{Y} \sim N_n(\boldsymbol{W\beta}, \sigma^2 \boldsymbol{I}_n)$. Therefore, $\hat{\beta}_3 \sim N(\beta_3, \sigma^2 a_{33})$ where $a_{33}$ is the corresponding diagonal element of $(\boldsymbol{W'W})^{-1}$.

Since the sampling distribution of $\hat{\beta}_3$ depends on the normality assumption of the errors and in small sample studies, this property may not hold, we will present a non-parametric approach to estimate the intervention effect on the trend using the Theil [11] and Sen[ 10] approaches. To estimate the effect, we split the data into pre- and post-intervention groups and calculate the slopes of each segment of the data. The procedure is as follows.

Let $y_{11}, y_{12}, \ldots, y_{1n_1}$ be the measurements of the response variable at $n_1$ points in time $t_{11}, t_{12}, \ldots t_{1n_1}$ for the pre-intervention time series. We then calculate $\frac{n_1(n_1-1)}{2}$ nonparametric slope estimates

$$\frac{y_{1j} - y_{1i}}{t_{1j} - t_{1i}} \tag{2.3}$$

for all $j > i$ and $i = 1,2,\ldots,(n_1 - 1)$ and $j = 2,3,\ldots,n_1$, where $y_{1j}$ and $y_{1j}$ are the measurements of the response variable at times $t_{1i}$ and $t_{1j}$ respectively for the pre-intervention segment of the data.

The estimate of trend (slope) for the pre-intervention segment of the data is $\hat{\beta}_{13} =$ median($\frac{y_{1j} - y_{1i}}{t_{1j} - t_{1i}}$) $\forall$ j>i . Similarly, for the post-intervention segment of data, the estimate of the trend (slope) is $\hat{\beta}_{23} =$ median($\frac{y_{2j} - y_{2i}}{t_{2j} - t_{2i}}$) $\forall$ j>i where $i = 1,2, \ldots, (n_2 - 1)$ and $j = 2,3, \ldots, n_2$.

Then the intervention effect on the trend is calculated as follows:

$$\hat{\beta}_3 = \hat{\beta}_{23} - \hat{\beta}_{13}. \tag{2.4}$$

## 2.2 Hypothesis testing:

If the intervention effect on the trend does not exist, the difference in slopes, which is $\beta_3$, has to be zero. Therefore, the test for no effect on the trend is:

$$H_0: \beta_3 = 0.$$

Testing $H_0: \beta_3 = 0$ based on the normality assumption is outlined below. The test statistic under the null hypothesis is

$$\frac{\hat{\beta}_3}{\sqrt{a_{33}\hat{\sigma}^2}} \sim t_{n-4}$$

where $a_{33}$ is the corresponding diagonal element of $(W'W)^{-1}$, $\hat{\sigma}^2 = \frac{Y'Y - \hat{\beta}'W'Y}{n-4}$ and $\hat{\beta}_3$ is the OLS estimate of the intervention effect $\beta_3$ in (2).

However, the literature has shown that this test is not applicable for small sample sizes where the errors do not follow a normal distribution. Hence, we will propose a rank-based nonparametric test procedure to test the hypothesis $H_0: \beta_3 = 0$. Based on the intervention effect calculated in (2.4), testing $H_0: \beta_3 = 0$ is equivalent to the two-sample median test.

Let the slopes calculated in (2.3) for both the pre-intervention and post-intervention data be two mutually independent random samples $V_1, V_2, \ldots, V_{N_1}$ and $W_1, W_2, \ldots, W_{N_2}$ from populations with continuous cumulative distribution functions $F$ and $G$ respectively.

The null hypothesis $H_0: \beta_3 = 0$ is equivalent to $H_0: F(t) = G(t)$ for all $t$ against the one-sided alternative hypothesis that W is larger (or smaller) than V. The alternative hypothesis can be written as $H_a: G(t) = F(t - \beta_3)$, for all $t$ where $\beta_3$ is the shifted amount (intervention effect). If $\beta_3 = 0$, the intervention effect is zero. If $\beta_3 > 0$, the intervention effect is positive, and if $\beta_3 < 0$, the effect is negative. In the location-shift model, the null hypothesis reduces to $H_0: \beta_3 = 0$ against $H_a: \beta_3 \neq 0$.

In literature, many nonparametric two-sample tests have been proposed such as the Wilcoxon rank sum test, the Mann-Whitney test [20], the Wald-Wolfowitz run test, and the Kolmogorov-Smirnov two sample test, etc. . However, to test $H_0: \beta_3 = 0$ against $H_a: \beta_3 \neq 0$, we here consider a bootstrap testing procedure. The reason for this is that we need to know the sampling distribution of the test statistic under the null hypothesis $H_0: \beta_3 = 0$, but deriving the sampling distribution of the intervention effect, $\hat{\beta}_3$, defined in (2.4) analytically is intractable. The bootstrap procedure for a two-sample test statistic for $H_0: \beta_3 = 0$ will be derived below.

Bootstrapping [7, 8, 9] is a computer-intensive approach to statistical inference. The idea behind bootstrapping is that sample information is used as a "proxy population". One takes samples with replacement from the original sample and calculates the statistic of interest repeatedly. The bootstrap method is easy to understand and implement, and it does not require any normality assumption of the sample data [9, 12]. For more details about bootstrapping see [9, 13].

Under the null hypothesis, the pre-intervention distribution of the slope, $F(v)$, is statistically the same as the post-intervention distribution of the slope, $G(w)$. We draw B bootstrap samples of size $N_1 + N_2$ from the same distribution, and each bootstrap sample represents the combined samples of $V_1, V_2, \ldots, V_{N_1}$ and $W_1, W_2, \ldots, W_{N_2}$. Let the first $N_1$ observations be denoted as $V^*$ and the remaining $N_2$ observations denoted as $W^*$. The intervention effect $\hat{\beta}_3^{*b} = \hat{\beta}_{23}^{*b} - \hat{\beta}_{13}^{*b}$, $b = 1,2, \ldots, B$ is then calculated for each bootstrap sample.

The approximate achieved significance level (ASL) is $\frac{\#\{|\hat{\beta}_3^{*b}| \geq |\hat{\beta}_3^0|\}}{B}$, where $\hat{\beta}_3^0$ is the observed intervention effect. If there is no intervention effect, ASL should be close to 0.5. If ASL is less than the significance level $\alpha = 0.05$, we reject the null hypothesis $H_0: \beta_3 = 0$.

To construct the percentile bootstrap confidence interval for $\beta_3$, for each bootstrap sample, we calculate $\hat{\beta}_3^{*b} = \hat{\beta}_{23}^{*b} - \hat{\beta}_{13}^{*b}$, $b = 1,2, \ldots, B$ and these resulting $\hat{\beta}_3^{*b}$'s are placed in ascending order. The $100(\frac{\alpha}{2})$th and $100(1 - \frac{\alpha}{2})$th percentile values of $\hat{\beta}_3^{*b}$ are selected resulting in a $100(1 - \alpha)$ percent percentile bootstrap confidence for $\beta_3$ of $(\hat{\beta}_3^{*b}{}_{\frac{\alpha}{2}}, \hat{\beta}_3^{*b}{}_{(1-\frac{\alpha}{2})})$.

## 3. Power comparison of two testing procedures:

In this section, we will compare the power of the two tests described above to test the null hypothesis $H_0: \beta_3 = 0$. The power of the test is the probability of rejecting the null hypothesis when it is false. The power for the t-test, and the two-sample median bootstrap test, described in the Section 2, will be estimated by Monte Carlo simulation under both standard normal and non-normal distributions for the errors.

Power = $\frac{\# \text{ rejected null hypotheses}}{R}$, where $R$ is the total number of simulation experiments.

The simulation was performed to generate 1000 independent and identically distributed random samples of the response variable $Y_t$ of size 16 using the model (2.1) with the error distribution: (I) exponential distribution with rate 0.10, and (II) standard normal, and the following design matrix:

Table 1: Design matrix for the segmented regression model

| $X_0$ | $T_t$ | $X_t$ | $X_t.(T_t - 8)$ |
|---|---|---|---|
| 1 | 1 | 0 | 0 |
| 1 | 2 | 0 | 0 |
| . | . | . | 0 |
| . | . | . | |
| . | . | . | |
| 1 | 8 | 0 | 0 |
| 1 | 9 | 1 | 1 |
| 1 | 10 | 1 | 2 |
| . | . | . | . |
| . | . | . | . |
| . | . | . | . |
| 1 | 16 | 1 | 8 |

The null regression coefficients configuration is $(\beta_0, \beta_1, \beta_2, \beta_3)$ =(4,4,0,0). The effect size for $\beta_3$ is set to be (1, 2,…,10) and (0.1, 0.2,…,0.9, 1.0) when the error distributions are exponential and standard normal respectively. Statistical Computing package R and Statistical software STATA 15.0 are used to compute the power of the tests and confidence intervals for the intervention effect parameters, and to analyze the real data used in this article.

Table 2: Power of the tests under non-normality (Error distribution is exponential with rate 0.10).

| $\beta_3$ | 0 | 1 | 2 | 3 | 4 | 5 | 6 | 7 | 8 | 9 | 10 |
|---|---|---|---|---|---|---|---|---|---|---|---|
| t-test | 0.05 | 0.06 | 0.18 | 0.31 | 0.47 | 0.60 | 0.75 | 0.83 | 0.90 | 0.94 | 0.97 |
| Robust median test | 0.05 | 0.08 | 0.20 | 0.36 | 0.62 | 0.82 | 0.93 | 0.97 | 0.99 | 0.99 | 1.00 |

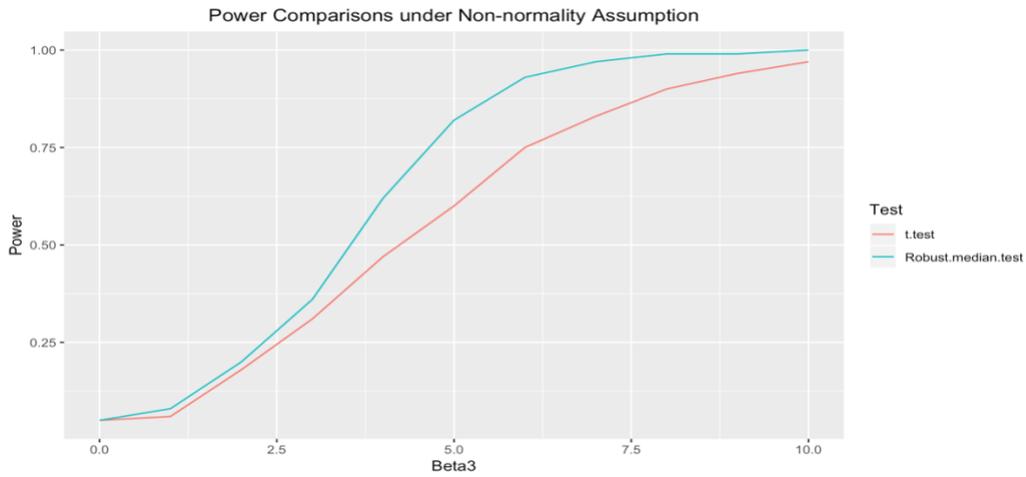

Figure1: Power comparisons under non-normality error distribution

Table 3: Power of the tests under normality (Error distribution is standard normal).

| $\beta_3$ | 0 | 0.1 | 0.2 | 0.3 | 0.4 | 0.5 | 0.6 | 0.7 | 0.8 | 0.9 | 1.0 |
|---|---|---|---|---|---|---|---|---|---|---|---|
| t-test | 0.05 | 0.07 | 0.14 | 0.24 | 0.39 | 0.56 | 0.71 | 0.84 | 0.92 | 0.97 | 0.99 |
| Robust median test | 0.05 | 0.06 | 0.12 | 0.21 | 0.36 | 0.54 | 0.69 | 0.83 | 0.91 | 0.96 | 0.98 |

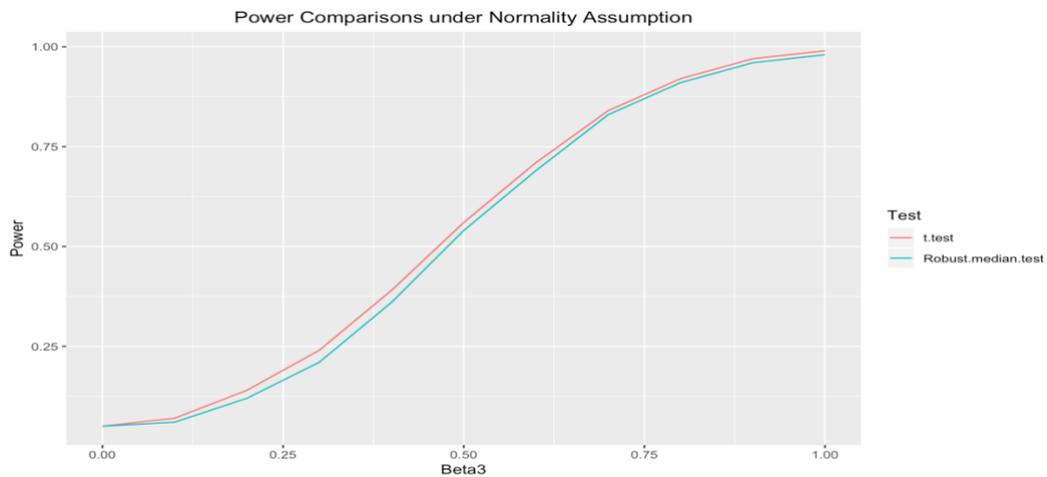

Figure2: Power comparisons under normality error distribution

In segmented regression analysis, to use the t-test for statistical inference regarding the regression coefficients, we must assume the residuals are normally distributed. If this assumption does not hold but we have large sample sizes, we can use an approximate test. However, with small sample sizes, and a severe departure from normality, the normal approximation results in low power of rejecting the correct alternative hypothesis (incorrectly declares no intervention effect on the trend). As shown in figure 1 above, the power of the robust median test is higher than the standard t -test under the non-normality and a small sample size of 16.

On the other hand, under the normality assumption of the error distribution, the standard method produces a slightly higher power of rejecting the incorrect null hypothesis (i.e. correctly identifies the intervention effect on the trend) than the robust median test does. However, the improvement in power using the standard method is very small, and even using small sample sizes, both methods are essentially equally effective in identifying the intervention effect on the trend (see figure 2 above). Furthermore, both methods are very sensitive in detecting small changes in the trend under the normality assumption of the error distribution.

4. Applications:

Organ transplantation is the most effective method of treating terminal organ failure, both in developed and developing countries [16, 19]. However, organ trafficking, transplant tourism, and transplant commercialism threaten to undermine the safe practice of transplantation worldwide [17].

The Istanbul Declaration 2008 outlines specific actions that take measures to protect the poorest and most vulnerable groups from transplant tourism and the sale of tissues and organs, as well as ways to reduce the prevalence of international trafficking in human tissues and organs [19, 21]. Within the Istanbul Declaration framework, the signee country must introduce programs to control and reduce the unethical practices of organ trafficking. However, numerous reports indicate that organ trafficking and transplant tourism continue to increase even though more than 100 countries have signed to control these unethical practices. In this section, we apply the two procedures described earlier to evaluate the impact of the Istanbul Declaration 2008 on the reduction of illegal and immoral practices on human organ trafficking.

We collected incidences of organ transplantation-related crimes reported on the Internet from 11 randomly selected countries from among the original participants of the Declaration. Selected countries include Brazil, Columbia, Egypt, India, Iran, Mexico, Nigeria, Pakistan, Philippines, Thailand, and Turkey. Incidence rates were collected from 2002-2015 for each country as reported in 2018. The population of each country was obtained from the World Bank database [18]. The occurrence rate per million population for each country was then calculated and used as the response.

We split the data into two categories; the pre-intervention time series 2002-2007, and the post-intervention time series 2008-2015 based on the year of the Declaration as an intervention. We calculated the change in the trend using two methods.

The time series plot in Figure 3 below shows a clear upward trend in organ trafficking for the period 2002 to 2015. In some countries, the increase in organ trafficking seems to increase over time leading to a non-linear trend. However, in other countries, the upward trend is linear and not as dramatic.

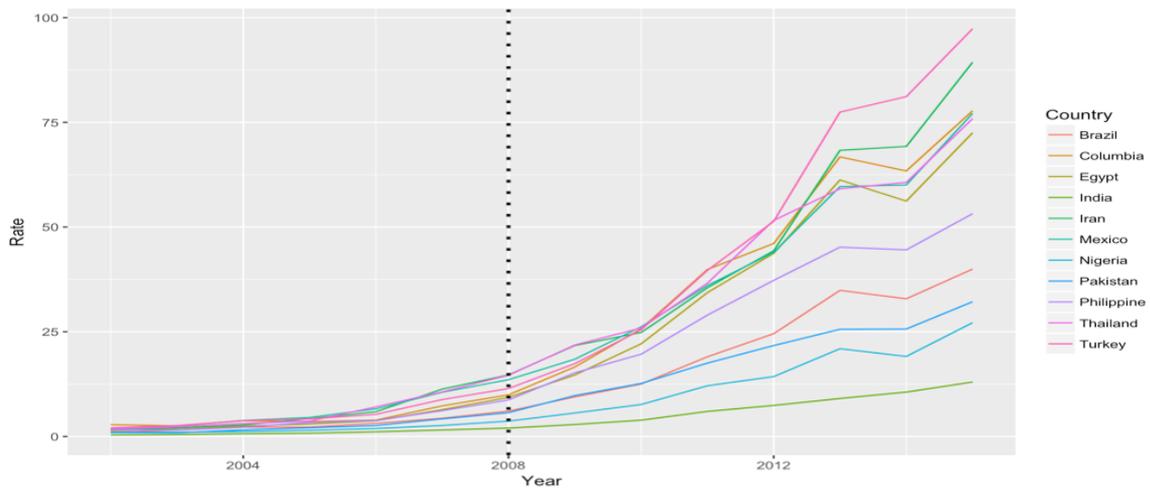

Figure 3: Time series plot for the rate of organ trafficking per million population

Recall that one of the assumptions for segmented regression is that there is a linear relationship between the outcome variable and time. To that end, we calculated the logarithm of the rate per million population to make the data linear.

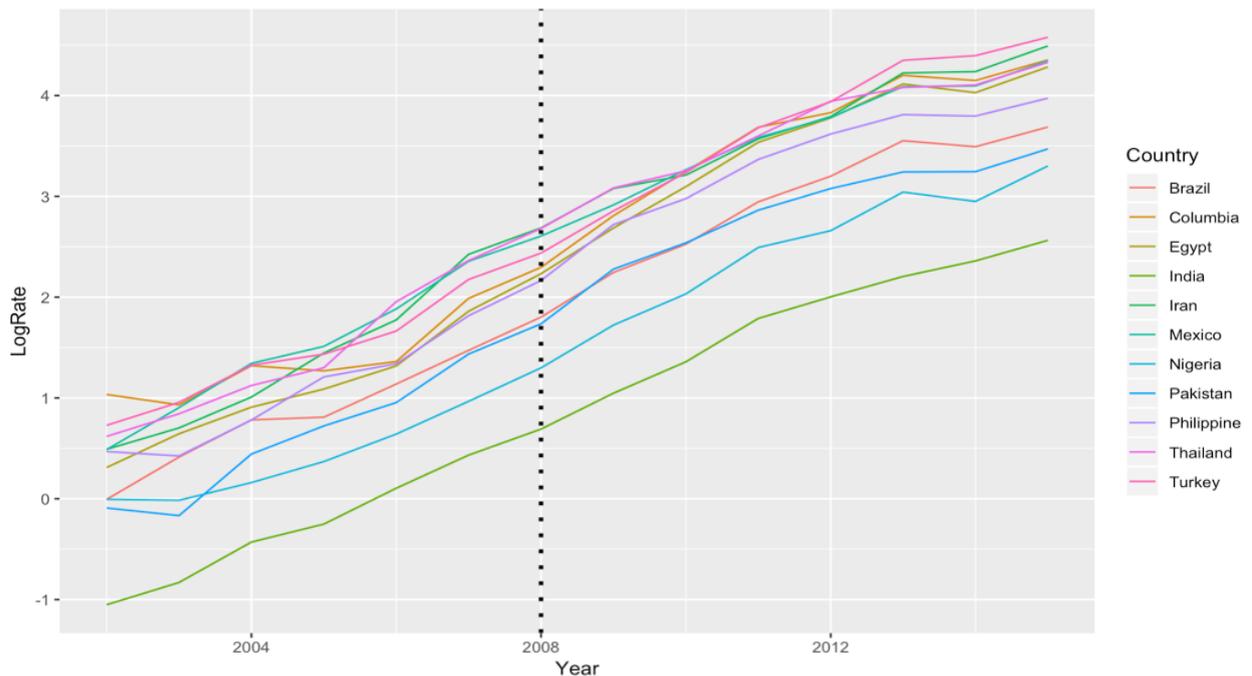

Figure 4: Time series plot for the Lograte of organ trafficking per million population

Since the presence of autocorrelation causes the standard errors of the estimates to be inflated, and consequently the confidence interval becomes wider, it is likely to come up with the wrong conclusion about the intervention effect. To detect the autocorrelation in the measurements, we use the Durbin Watson test. If it exists in the data, we control the autocorrelation based on NEWEY-WEST [14] and PRAIS [15] methods.

Table 4: Standard Method Estimation of the Intervention Effect on Linear Trend ($\beta_3$).

| Country | $\hat{\beta}_3$ | $\widehat{se}(\hat{\beta}_3)$ | 95% CI for $\beta_3$ |
|---|---|---|---|
| Brazil | -0.04 | 0.0272 | (-0.09, 0.02) |
| Colombia | 0.04 | 0.0544 | (-0.07, 0.14) |
| Egypt | -0.05 | 0.0296 | (-0.11, 0.01) |
| India | -0.05** | 0.0205 | (-0.09, -0.01) |
| Iran | -0.01 | 0.0214 | (-0.06, 0.03) |
| Mexico | -0.01 | 0.0208 | (-0.05, 0.03) |
| Nigeria | 0.03 | 0.0382 | (-0.05, 0.10) |
| Pakistan | -0.14** | 0.0358 | (-0.21, -0.07) |
| Philippines | -0.09** | 0.0398 | (-0.17, -0.01) |
| Thailand | -0.15** | 0.0268 | (-0.20, -0.10) |
| Turkey | 0.01 | 0.0256 | (-0.04, 0.06) |

** statistically significant results

Table 4 above reports estimates of the linear trend effects, their associated standard errors, and 95% confidence intervals, for each country. These estimates were calculated using the standard technique that assumes normally distributed errors. We see that for all countries in our sample except Columbia, Nigeria and Turkey ($\hat{\beta}_3$ = 0.04, 0.03, and 0.01 respectively) the differences in pre-intervention and post-intervention regression slopes are negative. But what does this tell us about the incidence of organ trafficking since the Istanbul Declaration?  Some care should be given to the interpretation of $\hat{\beta}_3$ since it is being used to predict $\ln(Y_t)$ and not $Y_t$ itself.

First consider the pre-intervention model defined as:

$$\ln(Y_t) = \beta_0 + \beta_1 T_t + \varepsilon_t. \qquad (4.1)$$

How do we interpret $\beta_1$ in the context of the original organ trafficking rate, $Y_t$? It is best to think in terms of *percent changes* of $Y_t$. It can be shown [22] that for every unit increase in $T_t$ (i.e. from one year to the next) the *percent change* in $Y_t$ is

$$(e^{\beta_1} - 1) \times 100 \approx 100\beta_1. \qquad (4.2)$$

The model defined in equation (4.1) indicates that the *percent increase* in organ trafficking incidents will be constant from one year to the next, which means that the *actual increase* in organ trafficking incidents will be larger from one year to the next. This is clearly seen in the plot of the original time series data in Figure 3. So, what does this mean for $\beta_3$? The post-intervention model is defined as:

$$\ln(Y_t) = (\beta_0 + \beta_2) + (\beta_1 + \beta_3)T_t + \varepsilon_t. \qquad (4.3)$$

Post-intervention, or since the Istanbul Declaration 2008, the *percent change* in $Y_t$ is now:

$$\left(e^{\beta_1+\beta_3} - 1\right) \times 100 \approx 100(\beta_1 + \beta_3). \tag{4.4}$$

Consider the example where $\beta_1 = .05$ and $\beta_3 = -.01$. This would indicate that the constant percent change in $Y_t$ is approximately 5% pre-intervention, and approximately 4% post-intervention. Researchers would conclude that organ trafficking is still increasing since the Istanbul Declaration, but at a slower rate. Returning to the results from Table 3, this is the exact conclusion regarding organ trafficking in all but three countries. However, this decline in the rate of reported 'organ trafficking' was statistically significant in only four countries: India (95% CI: -0.09 to -0.01), Pakistan (95% CI: -0.21 to -0.07), Philippines (95% CI: -0.17,-0.01), and Thailand (95% CI: -0.20 to -0.10).

Table 5 below reports the results for the robust method. While the signs of the estimated trend effects are all the same, there are noticeable differences in the magnitude of these coefficients, as well as the number of statistically significant results. The estimated reduction in trends for Iran and Mexico went from -0.01 using the standard method to -0.145 and -0.115 respectively using the robust method. Furthermore, the estimated reduction in trend was statistically significant in six countries (India, Iran, Mexico, Pakistan, Philippines, and Thailand) using the robust method instead of only three. Finally, the length of a 95% confidence interval for $\beta_3$ is slightly shorter using the robust median method rather than the standard method, implying that the robust method is more accurate as well.

Table 5: Robust Method Estimation of the Intervention Effect on Linear Trend ($\beta_3$)

| Country | $\widehat{\beta}_3$ | p-value | 95% CI for $\beta_3$ |
|---|---|---|---|
| Brazil | -0.046 | 0.245 | (-0.08, 0.02) |
| Colombia | 0.047 | 0.417 | (-0.05, 0.14) |
| Egypt | -0.044 | 0.465 | (-0.08, 0.04) |
| India | -0.052** | 0.006 | (-0.10, -0.02) |
| Iran | -0.145** | 0.001 | (-0.18, -0.09) |
| Mexico | -0.115** | 0.001 | (-0.16, -0.07) |
| Nigeria | 0.002 | 0.572 | (-0.04, 0.10) |
| Pakistan | -0.130** | 0.001 | (-0.18, -0.07) |
| Philippines | -0.110** | 0.002 | (-0.17, -0.04) |
| Thailand | -0.150** | 0.001 | (-0.20,-0.08) |
| Turkey | 0.006 | 0.658 | (-0.04, 0.09) |

** statistically significant results

5. Conclusions:

Segmented regression is a very powerful technique to detect changes in trend after the intervention on time series data. However, researchers should exercise caution when faced with small sample sizes. The standard regression technique includes the assumptions of linearity between time and the outcome variable, statistical independence (no autocorrelation) of the errors, homoscedasticity of the errors, and normality of the error distribution [1,3,6]. Transformations can be made in the presence of non-linearity and heteroscedasticity, and adjustments can be made for autocorrelation as well (NEWEY-WEST [14] and PRAIS [15]). However, the normality assumption is not robust for small sample sizes. If researchers are faced with non-normality and small sample sizes, the robust nonparametric median test will provide more power in significance testing, and more precise confidence intervals in estimation. This was demonstrated both by simulation, and when applied to time series data on organ trafficking, before and after the Istanbul Declaration.